\documentclass[aps,12pt,superscriptaddress,amsfonts,amssymb,amsmath,nofootinbib]{revtex4}

\usepackage{xcolor}

\usepackage{epsfig}
\usepackage{makeidx}
\usepackage{graphicx}
\usepackage{epstopdf}

\begin{document}

\def\pb{\, .}
\def\vb{\, ,}

\title{Microscopic Models for Welfare Measures\\ 
        addressing a Reduction of Economic Inequality}

\author{Maria Letizia Bertotti \footnote{Email address: marialetizia.bertotti@unibz.it}}
\author{Giovanni Modanese \footnote{Email address: giovanni.modanese@unibz.it}}
\affiliation{Free University of Bozen-Bolzano\\
                   Faculty of Science and Technology\\
                   39100 Bolzano, Italy}

\linespread{0.9}

\bigskip

\begin{abstract}

\ 
\ 
\ 

We formulate a flexible micro-to-macro kinetic model which is able to explain the emergence of income profiles out of a whole of individual economic interactions. 
The model is expressed by a system of several nonlinear differential equations which involve parameters defined by probabilities. 
Society is described as an ensemble of individuals divided into income classes; the individuals exchange money through binary and ternary interactions, 
leaving the total wealth unchanged. The ternary interactions represent taxation and redistribution effects. Dynamics is investigated through computational simulations, 
the focus being on the effects that different fiscal policies and differently weighted welfare policies have on the long-run income distributions. 
The model provides a tool which may contribute to the identification of the most effective actions towards a reduction of economic inequality. 
We find for instance that, under certain hypotheses, the Gini index is more affected by a policy of reduction of the welfare and subsidies for the rich classes 
than by an increase of the upper tax rate. Such a policy also has the effect of slightly increasing the total tax revenue.

\smallskip

\end{abstract}

\maketitle




\section{Introduction}
\label{Intro}
Economic inequality among individuals is a longstanding phenomenon which affects, to a large or small degree, most countries. 
A certain amount of inequality is unavoidable in a free market economy, especially at times of strong growth. 
High inequality levels, however, have been recognized to be harmful for economic development, 
to be one of the typical causes of political instability, and to be often at the origin of several social problems, 
including mental and physical diseases (see e.g. \cite{Deaton,Persson T. Tabellini G.,WilPic}).
Tackling inequality is not an easy task and is usually supposed to be a matter for economists, sociologists and politicians. 
The most effective actions for keeping it under control are probably provided by a proper fiscal system and by suitably targeted welfare policies.

We think that, to some extent, mathematics can also give a contribution in this direction. 
Through conceptual models and numerical simulations it can help to recognize and understand the mechanisms 
which cause inequality in the first place, and then let it persist. Mathematical models allow the exploration 
of different conceivable scenarios and, supported by real world data, may even suggest suitable concrete policies.

In this paper, without any inappropriate ambition to provide solutions, we formulate a set of simplified models 
for the description of the micro-processes of money exchange, taxation and redistribution in a closed market society. 
We show the emergence, from these micro-processes, of collective patterns like the income distribution curve. 
Our models are quite flexible and allow to consider different fiscal systems, characterized, for instance, 
by different tax rates for the income classes and by different gaps between the maximum and minimum tax rate. 
In addition, the models include the possibility that welfare provisions are specifically weighted for each income class. 
The focus in this paper is precisely on the effect of variable taxation and welfare on the income distribution and on its unevenness.
 
While traditional treatments of these subjects in mainstream economics rely on the assumption of a representative rational agent, 
our approach fits in with a complex system perspective. Arguments in favour of such a perspective can be found e.g. in
\cite{Arthur B. Durlauf S. Lane D.A.,Gallegati M. Kirman A.,Kirman A.,Gilbert N. Bullock S.,Squazzoni F.}.
We look at a population (or society) as a system composed by 
a large number of heterogeneous elements - the individuals - and we identify their interactions as the basic ingredients of the overall process. 
The observable collective features result from the interplay of these interactions. Accordingly, the system manifests self-organization. 
Due to its analytical character, our approach also differs from others which adopt the same interaction-based perspective. 
The tools most frequently employed in the study of 
socio-economic
complex systems are indeed agent-based models,
often also in combination with
a complex network structure
(see e.g. 
\cite{Tesfatsion L.,Tesfatsion L. Judd K.L.}).
Our models 
are expressed by systems of nonlinear ordinary differential equations of the kinetic-discretized Boltzmann type. 
The differential equations are as many as the classes, distinguished by their average income, in which one divides
the population. 
The $j$-th equation (with $j = 1, ..., n$) describes the variation in time of the fraction $x_{j}$ of individuals belonging
to the $j$-th class, and the modellization of this variation involves 
stochastic 
elements,
represented by
the presence in the equations
of suitable transition probabilities.
These models constitute an evolution of models originally introduced in \cite{Bertotti M.L.}
and then variously modified (so as to allow the study of different questions) and
investigated in \cite{Bertotti M.L. Modanese G. 1,Bertotti M.L. Modanese G. 2,Bertotti M.L. Modanese G. 3,Bertotti M.L. Modanese G. 4}.
The main novelty of the present paper is given by
the inclusion in the models of differently weighted welfare measures,
aimed at recognizing  ways to prevent an excessive inequality.

In a way, our models belong to the wide class of Asset Exchange Models \cite{Krapivsky P.L. Redner S.}, sometimes also called ``Yard Sale" models 
because of their simplified representation of economic interactions. The main purpose of this representation is to
emphasize the statistical consequences of the interactions of a large number of agents. It is well known in statistical mechanics that certain aggregate features of large ensembles do not depend of the details of the interactions,
but emerge as general consequences of statistics. In view of the universal features displayed by income distributions in several countries and in different epochs, it has long been suspected that these features arise from some
common general mechanism. Asset Exchange Models could then be regarded as the economical equivalent of the perfect gas model or the Carnot engine in physics.
This position has recently been advocated by Boghosian \cite{Boghosian B.M.}, who proposed a continuum 
model and derived, in a suitable limit, a 
partial differential equation governing its evolution. 
An improvement of the model, which was explored in \cite{Boghosian B.M.}
and is not yet present in our approach, is the introduction of economic ``production", in such a way that the total
income of the society is not constant in time.
We observe that the redistribution terms considered in \cite{Boghosian B.M.} correspond to an income tax independent from the single interactions, while in our scheme the taxation is 
both related to the income and applied to each single transaction, and therefore 
also includes a valued added tax component.

There are further reasons to believe that kinetic models capture the ``game-theoretical" strategies of the interacting agents much more than their physical origin could suggest. 
In fact, it was shown in \cite{Challet D. Marsili M. Zecchina R.,Venkatasubramanian V.} for two different
game-theoretical models, that their large-numbers averages are equivalent to those of certain statistical mechanics systems. This suggests that some models of statistical mechanics originally developed to describe ensembles of
inanimated atoms or spins are suitable also for the description of ensembles of individuals following a strategy, because in both cases the system tends to minimize some objective function.

\smallskip

The paper is organized as follows. In the next section a model family
suitable for handling the issue from a complexity viewpoint is introduced.
Some key properties of the family models, and the emergence of the relative aggregate outcomes
together with their features resulting from several numerical simulations are investigated and discussed in the third section. 
In particular, comparisons between different models and different policies
are drawn. Finally, in the last section, a summary
with a critical analysis is given and further developments and perspectives are mentioned.

\section{A family of models encompassing different welfare measures for different income classes}
\subsection{A general framework}
The framework within which the model family can be constructed was first introduced in \cite{Bertotti M.L.}.
We refer to that paper for a detailed illustration of the stylized micro-scale mechanism 
that the framework aims to describe.

Imagine dividing a population of individuals into a finite number $n$ of classes, each one characterized by its average income,
the average incomes being the positive numbers $r_1< r_2 < ... <\ r_n$. 
Here, we just recall that also the part of the government (which of course plays a role in connection with the taxation system)
can be described through monetary exchanges between pairs of individuals,
and we emphasize that consequently two kinds of interactions may take place: the so called {\it direct} ones, between an 
$h$-individual and a $k$-individual, occurring when the first one pays the second one,
and the {\it indirect} ones, between the $h$-individual and every $j$-individual
with $j \ne n$, occurring on the occasion of the direct $h$-$k$ interaction. The indirect interactions
represent the transactions corresponding to the payment of taxes and to the benefit of the redistribution.
In short, and we are referring here to a tax compliance case, in correspondence to any direct $h$-$k$ interaction, 
if $S$ (with $S < (r_{i+1} - r_{i})$ for all $i = 1, ..., n$)
denotes the amount of money that the $h$-individual should pay to the $k$-one, the overall effect of payment, 
taxation and redistribution is that of an $h$-individual paying a quantity $S \, (1 - \tau)$ to a $k$-individual 
and paying as well a quantity $S \, \tau$, which is divided among all $j$-individuals 
for $j \ne n$.\footnote{\ Individuals of the $n$-th class cannot receive money. Otherwise, they would possibly
advance to a higher class. And this is not permitted in the present context.} The quantity $\tau = \tau_k$, which is 
assumed to depend on the class to which the earning individual belongs, corresponds to the taxation rate of the $k$-th class.

The taxation and redistribution processes relative to such a population
can be modelled within the framework provided by the system of $n$ nonlinear differential equations 
\begin{equation}
{{d x_i} \over {d t}} =  
\sum_{h=1}^n \sum_{k=1}^n {\Big (} C_{hk}^i + T_{[hk]}^i(x) {\Big )}
x_h x_k     -    x_i  \sum_{k=1}^n x_k \vb \qquad 
i= 1\vb ... \vb n \pb
\label{evolution eq eta = 1}
\end{equation}
Here, $x_i(t)$ with $x_i : {\bf R} \to [0,+\infty)$ denotes the fraction at time $t$
of individuals belonging to the $i$-th class; the coefficients $C_{hk}^i \in [0,+\infty)$, 
satisfying $\sum_{i=1}^n C_{hk}^i = 1$ for any fixed $h$ and $k$,
represent transition probability densities due to the direct interactions (more precisely, $C_{hk}^i $
expresses the probability density that an individual of the $h$-th class will belong to the 
$i$-th class after a direct interaction with an individual of the $k$-th class), and the functions
$T_{[hk]}^i : {\bf R}^n \to {\bf R}$, continuous and 
satisfying $\sum_{i=1}^n T_{[hk]}^i(x) = 0$ for any fixed $h$, $k$ and $x \in {\bf R}^n$, 
represent transition variation densities due to the indirect interactions
(more precisely, $T_{[hk]}^i$ expresses the 
variation density in the $i$-th class
due to an interaction between an individual of the $h$-th class
with an individual of the $k$-th class).
The system $(\ref{evolution eq eta = 1})$
accounts for the fact that any direct or indirect interaction
possibly causes a slight increase or decrease of the income of individuals.

\subsection{Construction of a specific model family}
\label{subsection:construction}
We start by defining certain coefficients $p_{h,k}$ for $h, k = 1, ... , n$,
which express the probability that  in an encounter between an $h$-individual and a $k$-individual,
the one who pays is the $h$-individual. Since also the possibility that none of the two pays has to be taken into account, the requirement 
which the $p_{h,k}$ must satisfy is that $0 \le p_{h,k} \le 1$ and $p_{h,k} + p_{k,h} \le 1$.
We take
$$
p_{h,k} = \min \{r_h,r_k\}/{4 r_n} \vb
$$
with the exception of the terms
$p_{j,j} = {r_j}/{2 r_n}$ for $j = 2, ..., n-1$,
$p_{h,1} = {r_1}/{2 r_n}$ for $h = 2, ..., n$, 
$p_{n,k} = {r_k}/{2 r_n}$ for $k = 1, ..., n-1$,
$p_{1,k} = 0$ for $k = 1, ..., n$
and
$p_{hn} = 0$ for $h = 1, ..., n$.
This choice, among others, was proposed and discussed in \cite{Bertotti M.L. Modanese G. 2}.
It gives account of a degree of heterogeneity among
individuals belonging to different classes, also in connection with their interactions with others.

We are now ready to construct a particular family of models. This will be done through the choice of
the values of the parameters $C_{hk}^i\in [0,+\infty)$ and of the functions $T_{[hk]}^i(x) : {\bf R}^n \to {\bf R}$.

\smallskip

As in \cite{Bertotti M.L.,Bertotti M.L. Modanese G. 1,Bertotti M.L. Modanese G. 2}, we assume that
the only possibly nonzero elements among the $C_{hk}^i$ are:
\begin{eqnarray}
C_{i+1,k}^{i} & = 
                  & p_{i+1,k} \, \frac{S \, (1-\tau_k) }{r_{i+1} - r_{i}} \vb \nonumber \\
C_{i,k}^i & = 
            & 1 - \, p_{k,i} \, \frac{S \, (1-\tau_i)}{r_{i+1} - r_{i}} 
               - \, p_{i,k} \, \frac{S \, (1-\tau_k)}{r_{i} - r_{i-1}} \vb \nonumber \\
C_{i-1,k}^i & = 
               & p_{k,i-1} \, \frac{S \, (1-\tau_{i-1})}{r_{i} - 
	       r_{i-1}} \pb 
\label{choiceforC}
\end{eqnarray} 
We stress that the expression for $C_{i+1,k}^{i}$ in $(\ref{choiceforC})$ holds true for $i \le n-1$ and $k\le n-1$;
the second addendum of the expression for $C_{i,k}^i$ is effectively present only 
provided $i \le n-1$ and $k \ge 2$, while its third addendum is present
only provided $i \ge 2$ and $k \le n-1$; the expression for $C_{i-1,k}^i$ holds true for $i \ge 2$ and $k\ge 2$.

\smallskip

As for the indirect transition variation densities $T_{[hk]}^i(x)$, we express them as
\begin{equation}
T_{[hk]}^i(x) =  
U_{[hk]}^i(x) + V_{[hk]}^i(x)
\vb
\label{T+U}
\end{equation}
where\footnote{\ 
In $(\ref{U_{[hk]}^i(x)})$,
$h >1$ 
and
the terms 
involving the index $i-1$ [respectively, $i+1$]
are effectively present only provided $i-1 \ge 1$ 
[respectively, $i+1 \le n$]. In other words:
the $1^o$ term into parentheses on the r.h.s. of (\ref{U_{[hk]}^i(x)}) is present for $2 \le i \le n$;
the $2^o$ term into parentheses on the r.h.s. of (\ref{U_{[hk]}^i(x)}) is present for $1 \le i \le n-1$.
}
\begin{equation}
U_{[hk]}^i(x) =  
\frac{p_{h,k} \, S \, \tau_k}{\sum_{j=1}^{n} w_j x_{j}} {\bigg (}  \frac{w_{i-1} x_{i-1}}{r_i - r_{i-1}} -   \frac{w_i x_{i}}{r_{i+1} - r_{i}} {\bigg )}
\label{U_{[hk]}^i(x)}
\end{equation}
represents
the variation density corresponding to the advancement
from a class to the subsequent one, due to the benefit of taxation
and\footnote{\ 
In $(\ref{V_{[hk]}^i(x)})$,
$h >1$ 
and
the terms 
involving the index $i-1$ [respectively, $i+1$]
are effectively present only provided $i-1 \ge 1$ 
[respectively, $i+1 \le n$]. In other words:
the $1^o$ term into parentheses on the r.h.s. of (\ref{V_{[hk]}^i(x)}) is present for $1 \le i \le n-1$;
the $2^o$ term into parentheses on the r.h.s. of (\ref{V_{[hk]}^i(x)}) is present for $2 \le i \le n$.
}
\begin{equation}
V_{[hk]}^i(x)
=  p_{h,k} \, S \, \tau_k \, \frac{\sum_{j=1}^{n-1} w_j x_{j}}{\sum_{j=1}^{n} w_j x_{j}} \, 
{\bigg (} 
\frac{\delta_{h,i+1}}{r_h - r_{i}} \, - \, \frac{\delta_{h,i}}{r_h - r_{i-1}}
{\bigg )} 
\vb
\label{V_{[hk]}^i(x)}
\end{equation}
with $\delta_{h,k}$ denoting the {\slshape Kronecker delta}, 
represents the variation density corresponding to the retrocession
from a class to the preceding one, due to the payment of some tax.

\noindent The coefficients $w_j$ in $(\ref{U_{[hk]}^i(x)})$ and $(\ref{V_{[hk]}^i(x)})$ denote the weights
here introduced to account 
for differently distributed welfare.
A conceivable expression for them is given e.g. by
\begin{equation}
w_j = r_{n+1-j} + \frac{2}{n-1} \, \gamma \, \bigg(j - \frac{n+1}{2}\bigg) \, (r_{n} - r_{1}) \ , 
\label{formula welfare}
\end{equation}
with $\gamma \in (0,1/2]$. The effect of
the parameter $\gamma$ is such that the smaller $\gamma$ is, the larger is the difference
$w_1 - w_n$. Indeed, $w_1 - w_n = (r_{n} - r_{1}) \, (1 - 2 \gamma)$.
And, if for example $r_j$ is taken to be linear in $j$,
$w_j$ decreases linearly as a function of $j$, with the exception of the limiting case when $\gamma = 1/2$,
in which $w_j$ has the same value for each $j = 1, ..., n$.

Notice that the effective amount of money representing taxes, which is paid 
in correspondence to a payment of $S (1 - \tau_k)$
and is then redistributed 
is $S \, \tau_k \,({\sum_{j=1}^{n-1} w_j x_{j}})/{(\sum_{j=1}^{n} w_j x_{j}})$
instead of $S \, \tau_k$.  This is a technical 
device, due to the bound on the income of individuals in the $n$-th class.

\medskip

To fix ideas, we take $S = 1$, 
\begin{equation}
r_j = 25 \, j \vb
\label{average incomes}
\end{equation}
\begin{equation}
\tau_j = \tau_{min} +  \frac{j - 1}{n-1} \, (\tau_{max} - \tau_{min}) \vb
\label{progressivetaxrates}
\end{equation}
for $j = 1, ... , n$, where $\tau_{max}$ and $\tau_{min}$ respectively denote the maximum and the minimum tax rate.
Still, the value of $\gamma$, $\tau_{min}$ and $\tau_{max}$
have to be fixed. 
Hence, the equations $(\ref{evolution  eq eta = 1})$ describe a family of models rather than a single model.
They are well beyond analytical solutions. But, relevant facts can be understood through simulations. 
To run simulations, we take here $n=15$.

\section{Properties of the model family}
We start here by stating three properties
which hold true for any model of the family introduced in the Subsection
$\rm{\ref{subsection:construction}}$.
While for the first two properties an analytical proof can be provided,
the third one is in fact only supported by a number of simulations.

\medskip

\noindent {\bf {Well posedness of the Cauchy problem}}.
For fixed values of the parameters $\gamma$, $\tau_{min}$ and $\tau_{max}$,
in correspondence to any initial condition $x_0 = (x_{01} , \ldots , x_{0n})$, 
for which $x_{0i} \ge 0$ for all $i = 1, ... , n$ and $\sum_{i=1}^n x_{0i} = 1$,
a unique solution $x(t) = (x_1(t),\ldots,x_n(t))$ of $(\ref{evolution eq eta = 1})$ exists,
which is defined for all $t \in [0,+\infty)$, satisfies $x(0) = x_0$ and also
\begin{equation}
x_{i}(t) \ge 0 \ \hbox{for} \ i = 1, ... , n \ \hbox{and} \ \sum_{i=1}^n x_{i}(t) = 1 \ \hbox{for all} \ t \ge 0 \, . 
\label{solution in the future}
\end{equation}
This has been analytically proved in \cite{Bertotti M.L.}
for similar models. 
The proof therein can be quite easily adapted to hold true in the present situation as well.

\medskip

\noindent {\bf {Conservation of the total income $\mu$}}.
For fixed values of the parameters $\gamma$, $\tau_{min}$ and $\tau_{max}$,
the scalar function
$\mu(x)=\sum_{i=1}^n r_i x_i$, expressing the global (and mean) income, 
is a first integral for the system $(\ref{evolution eq eta = 1})$. 
Again, the result can be proved by introducing some slight, obvious modifications in the 
corresponding proof 
in \cite{Bertotti M.L.}. 

\medskip

\noindent {\bf {Uniqueness, for any fixed value of $\mu$, of the asymptotic stationary distribution}}.
For fixed values of the parameters $\gamma$, $\tau_{min}$ and $\tau_{max}$,
for any fixed value $\mu \in [r_1,r_n]$, 
an equilibrium of $(\ref{evolution eq eta = 1})$ exists, to which
all solutions of $(\ref{evolution eq eta = 1})$,
whose initial conditions $x_0 = (x_{01} , \ldots , x_{0n})$ satisfy
$x_{0i} \ge 0$ for all $i = 1, ... , n$, $\sum_{i=1}^n x_{0i} = 1$, and $\sum_{i=1}^n r_i x_{0i} = \mu$
tend asymptotically as $t \to +\infty$.
In other words, a one-parameter family of asymptotic stationary distributions exists,
the parameter being the total income value. 

\medskip

Other properties are directly related to the issue of interest here. They concern comparisons
between models of the
family at hand, characterized by different
fiscal systems and different welfare policies.
To  
quantitatively evaluate the consequences of these differences,
it is useful to refer to indicators
as
the Gini index $G$ and the tax revenue $T \!R$.
We first recall here the definition of these two quantities.
Then we will try and see,
with reference to some specific examples,
which one among two conceivable policies is more efficient,
one of the two conceivable policies being the adoption
of a fiscal system characterized by a suitable spread between the maximum and the minimum income class tax rates
and the other one being
the introduction of suitably weighted welfare measures.

The definition of the Gini index involves the Lorenz curve, which plots the cumulative percentage of the total income 
of a population (on the $y$ axis) earned by the bottom percentage of individuals (on the $x$ axis).
Specifically, $G$ corresponds to the ratio $A_1/A_2$ of the area $A_1$ between the Lorenz curve and the line of perfect equality (the line at $45$ degrees)
and the total area $A_2$ under the line of perfect equality. It takes values in $[0,1]$, $0$ representing the complete equality and $1$ the maximal inequality. 

The tax revenue
is the total amount of tax collected in the unit time and redistributed as welfare provisions. 
It is given by
\begin{equation}
T\!R = \sum_{h=1}^{n} \, \sum_{k=1}^{n} \,\sum_{j=1}^{n - 1} \, \, p_{hk} \, \tau_k \, 
\frac{w_j {{\hat x}_j}}{(\sum_{i=1}^{n} w_i {{\hat x}_i})} \, {{\hat x}_h}{{\hat x}_k} \vb
\end{equation}
where ${\hat x}_i$ denotes the fraction of individuals in the $i$-th class at equilibrium.

\subsection{Adopting different fiscal systems}
Concerning the choice of different fiscal systems, we take into account the possibility of varying the tax rates,
expressed according to the progressive rule 
$(\ref{progressivetaxrates})$, by letting $\tau _{min}$ and $\tau_{max}$ change. In contrast, we fix here $\gamma = 0.5$, which amounts to grant the same 
welfare to each class.

The results for a couple of prototypical simulation sets are given next. In each of the two cases, 
a random initial income distribution is chosen, subject to the requirement that the majority of individuals belong to lower income classes.

\medskip

{\bf {Example}} $1$.
The value of the total income $\mu$ is here  equal to $\mu = 135.00$.
We consider different model versions, corresponding to different choices of the minimum and maximum tax rate 
$\tau _{min}$ and $\tau _{max}$. Letting the situation evolve, we then evaluate the Gini index relative to the asymptotic stationary distribution.
And we evaluate as well the tax revenue corresponding to this distribution.
The outputs are reported in Table $\rm{\ref{tab:1varyingtaxrates}}$. 

\begin{table}[h]
\begin{center}
\begin{tabular}{|c|c|c|c|}
\hline
$\tau _{min}$ \ & \ $\tau_{max}$ \ & \ Gini Index \ & \ Tax Revenue \nonumber \\
\hline
30\% \ & \ 45\%  \ & \ 0.368 \ & \ 0.0222 \nonumber \\
\hline
25\% \ & \ 50\% \ & \ 0.361 \ & \ 0.0219 \nonumber \\
\hline
20\% \ & \ 55\% \ & \  0.354 \ & \ 0.0215 \nonumber \\
\hline
15\% \ & \ 60\% \ & \  0.347 \ & \ 0.0210 \nonumber \\
\hline
10\% \ & \ 65\% \ & \ 0.341 \ & \ 0.0205 \nonumber \\
\hline
\end{tabular}
\end{center}
\caption
{The data in this table refer to models for which
different minimum and maximum tax rate are assumed.
The value of the Gini index and of the tax revenue relative to the asymptotic stationary distribution,
which are obtained through the numerical simulations, are given.
In all cases, the initial condition is the same. The value of the corresponding total income is here $\mu = 135.00$.}
\label{tab:1varyingtaxrates}
\end{table}

\medskip

{\bf {Example}} $2$.
As in the Example $1$ we
consider here different model versions, corresponding to different choices of the minimum and maximum tax rate 
$\tau _{min}$ and $\tau _{max}$.
The 
total income is $\mu = 127.65$.
The outputs are reported in Table $\rm{\ref{tab:2varyingtaxrates}}$. 

\begin{table}[h]
\begin{center}
\begin{tabular}{|c|c|c|c|}
\hline
$\tau _{min}$ \ & \ $\tau_{max}$ \ & \ Gini Index \ & \ Tax Revenue \nonumber \\
\hline
30\% \ & \ 45\%  \ & \ 0.378 \ & \ 0.0206 \nonumber \\
\hline
25\% \ & \ 50\% \ & \ 0.370 \ & \ 0.0201 \nonumber \\
\hline
20\% \ & \ 55\% \ & \  0.364 \ & \ 0.0196 \nonumber \\
\hline
15\% \ & \ 60\% \ & \  0.357 \ & \ 0.0190 \nonumber \\
\hline
10\% \ & \ 65\% \ & \ 0.350 \ & \ 0.0183 \nonumber \\
\hline
\end{tabular}
\end{center}
\caption
{The data in this table refer to models for which
different minimum and maximum tax rate are assumed.
The value of the Gini index and of the tax revenue relative to the asymptotic stationary distribution,
which are obtained through the numerical simulations, are given.
In all cases, the initial condition is the same. The value of the corresponding total income is here $\mu = 127.65$.}
\label{tab:2varyingtaxrates}
\end{table}

\medskip

The outputs of these and several other simulations can be summarized in the following statement. 

\smallskip

\noindent {\bf {Dependence of the asymptotic stationary distribution on $\tau_{min}$ and $\tau_{max}$}}.
The profile of the asymptotic stationary distribution depends on
$\tau_{min}$ and $\tau_{max}$.
For example, 
if the difference between the maximum and the minimum tax rates, $\tau_{max} - \tau_{min}$
is enlarged, while all other data are kept unchanged,
an increase of the fraction of individuals belonging to the middle classes 
(to the detriment of those in the poorest and richest classes)
can be detected at the asymptotic equilibrium.

\subsection{Adopting different welfare measures}
We then consider different model versions obtained incorporating a differentiated welfare system.
Specifically, we let the initial data of the Examples $1$ and $2$ evolve
in correspondence to systems for which the minimum and maximum tax rates are fixed (and amount in particular to $30 \%$ and $45\%$ respectively).
In addition to
the model in which the same welfare provision is guaranteed to each class,
we consider,
for both cases of Example $1$ and Example $2$,  
seven variants encompassing different welfare for different classes.
These variants are constructed through
the choice of different values, ranging from $0.45$ to $0.15$,
of the parameter $\gamma$.
Recall that 
providing the same welfare to each class amounts to
taking $\gamma = 0.5$.

The results we get are summarized in the Tables $\rm{\ref{tab:1diffwelfare}}$ and $\rm{\ref{tab:2diffwelfare}}$. 
In these tables, beside the values of the parameter $\gamma$, also the ratios $w_{15}/w_{1}$ are listed. They 
express the proportion of welfare which is granted to individuals of the richest class
with respect to that one granted to individuals of the poorest class.

\begin{table}[h]
\begin{center}
\begin{tabular}{|c|c|c|c|c|}
\hline
$\gamma$ \ & $w_{15}/w_{1}$ \ & \ Gini Index \ & \ Tax Revenue \nonumber \\
\hline
$0.50$ \ & \ 1 \ & \ 0.368 \ & \ 0.0222 \nonumber \\
\hline
$0.45$ \ & \ 0.84 \ & \ 0.363 \ & \ 0.0225 \nonumber \\
\hline
$0.40$ \ & \ 0.70 \ & \  0.358 \ & \ 0.0227 \nonumber \\
\hline
$0.35$ \ & \ 0.58 \ & \  0.353 \ & \ 0.0229 \nonumber \\
\hline
$0.30$ \ & \ 0.48 \ & \  0.349 \ & \ 0.0231 \nonumber \\
\hline
$0.25$ \ & \ 0.39 \ & \  0.345 \ & \ 0.0233 \nonumber \\
\hline
$0.20$ \ & \ 0.31 \ & \  0.341 \ & \ 0.0235 \nonumber \\
\hline
$0.15$ \ & \ 0.24 \ & \  0.338 \ & \ 0.0236 \nonumber \\
\hline
\end{tabular}
\end{center}
\caption
{The data in this table refer to models characterized by differentiated welfare policies. 
In all cases, the initial condition is the same: it is the same as in the Example $1$.}
\label{tab:1diffwelfare}
\end{table}

\begin{table}[h]
\begin{center}
\begin{tabular}{|c|c|c|c|c|}
\hline
$\gamma$ \ & \ $w_{15}/w_{1}$ \ & \ Gini Index \ & \ Tax Revenue \nonumber \\
\hline
$0.50$ \ & \ 1 \ & \ 0.378 \ & \ 0.0206 \nonumber \\
\hline
$0.45$ \ & \ 0.84 \ & \ 0.372 \ & \ 0.0208 \nonumber \\
\hline
$0.40$ \ & \ 0.70 \ & \  0.368 \ & \ 0.0210 \nonumber \\
\hline
$0.35$ \ & \ 0.58 \ & \  0.363 \ & \ 0.0212 \nonumber \\
\hline
$0.30$ \ & \ 0.48 \ & \  0.359 \ & \ 0.0214 \nonumber \\
\hline
$0.25$ \ & \ 0.39 \ & \  0.355 \ & \ 0.0215 \nonumber \\
\hline
$0.20$ \ & \ 0.31 \ & \  0.352 \ & \ 0.0217 \nonumber \\
\hline
$0.15$ \ & \ 0.24 \ & \  0.348 \ & \ 0.0218 \nonumber \\
\hline
\end{tabular}
\end{center}
\caption
{The data in this table refer to models characterized by 
differentiated welfare policies. 
In all cases, the initial condition is the same: it is the same as in the Example $2$.}
\label{tab:2diffwelfare}
\end{table}

Figures $1$ and $2$ give a visual representation. 
The two four-panel-blocks concern asymptotic 
solutions evolving from the same initial data 
considered in the Examples $1$ and $2$ respectively.
In each such block, the panels in the first row display 
the asymptotic stationary distribution of the model in which the welfare is given by the formula
$(\ref{formula welfare})$ 
with $\gamma = 0.45$ (on the left)
and 
with $\gamma = 0.15$ (on the right). As for the histograms in the second row:
that on the left
depicts for each class
the difference between the fraction of individuals pertaining to 
the model with $\gamma = 0.45$ and the fraction of individuals pertaining to 
the model with the same welfare for all classes;
that on the right depicts for each class
the difference between the fraction of individuals pertaining to 
the model with $\gamma = 0.15$ and the fraction of individuals pertaining to 
the model with the same welfare for all classes.

\begin{figure*}[tbp]
\begin{center}
\includegraphics[width=4.75cm,height=2.5cm] {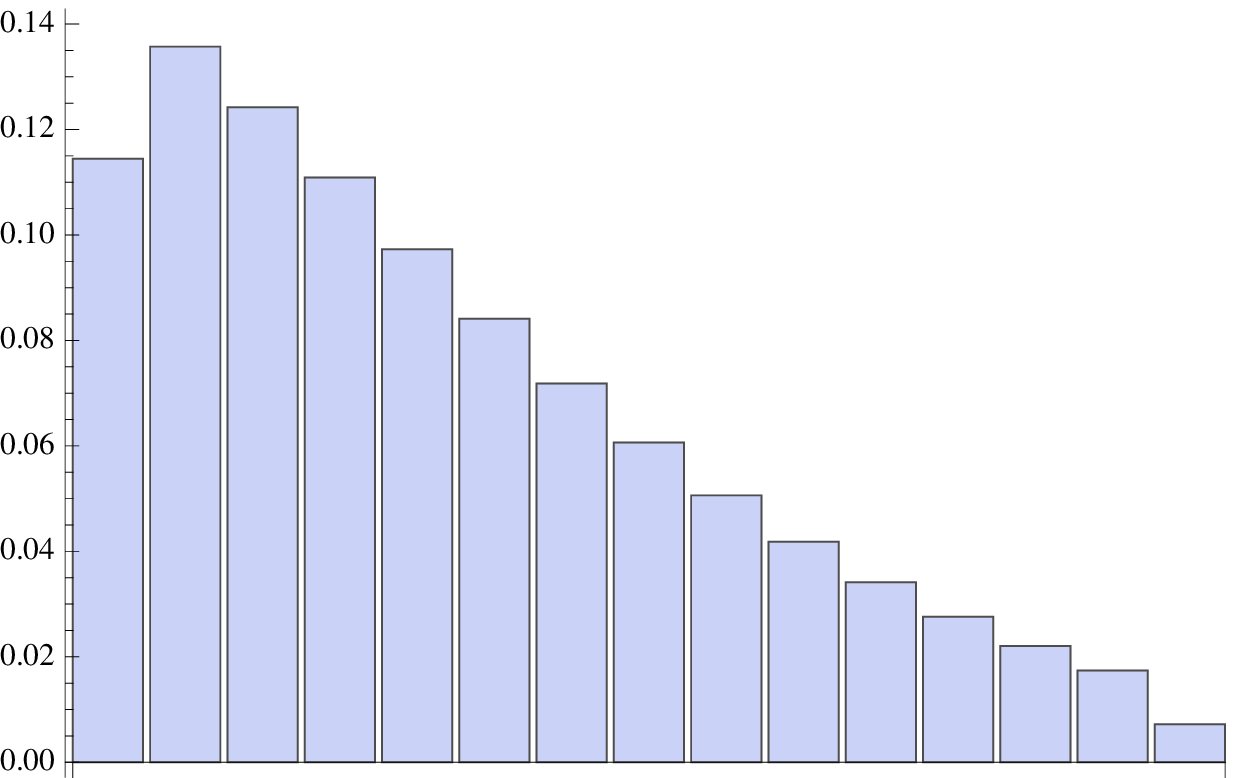}
 \hskip0.7cm
\includegraphics[width=4.75cm,height=2.5cm] {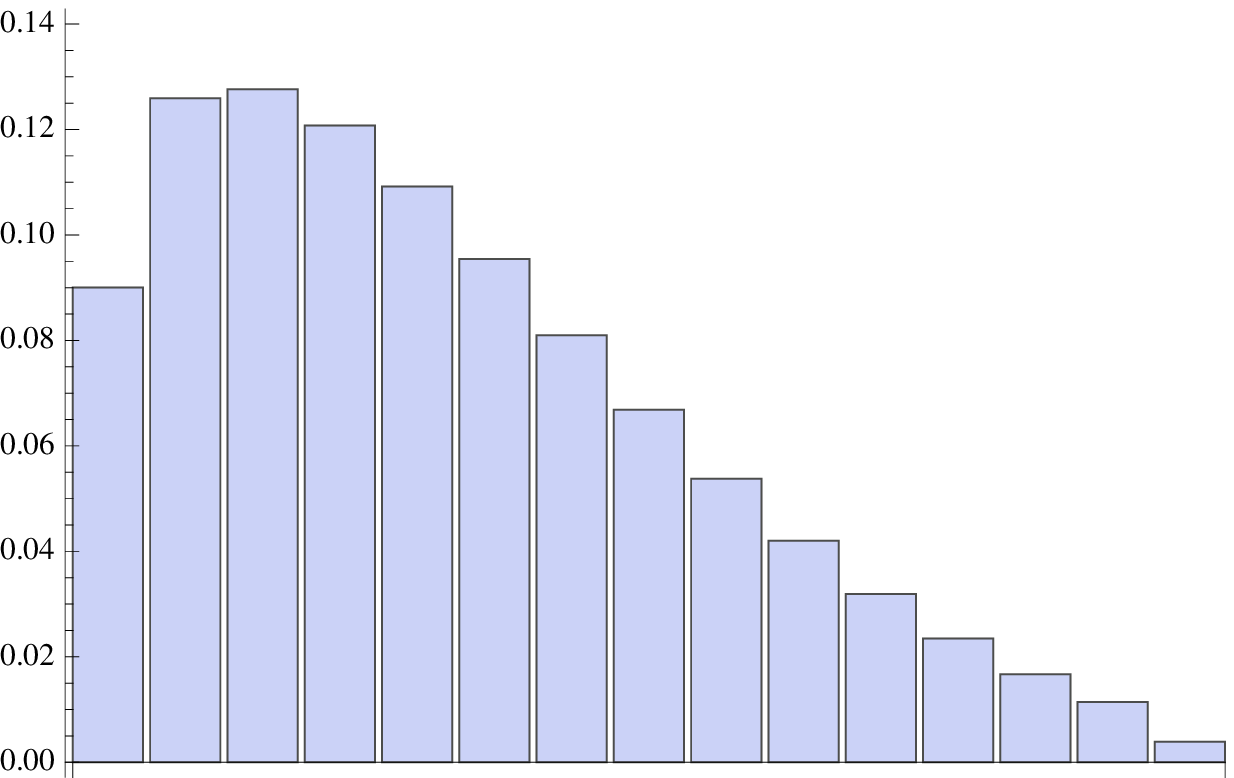}
\end{center}
\begin{center}
\includegraphics[width=4.75cm,height=1.55cm] {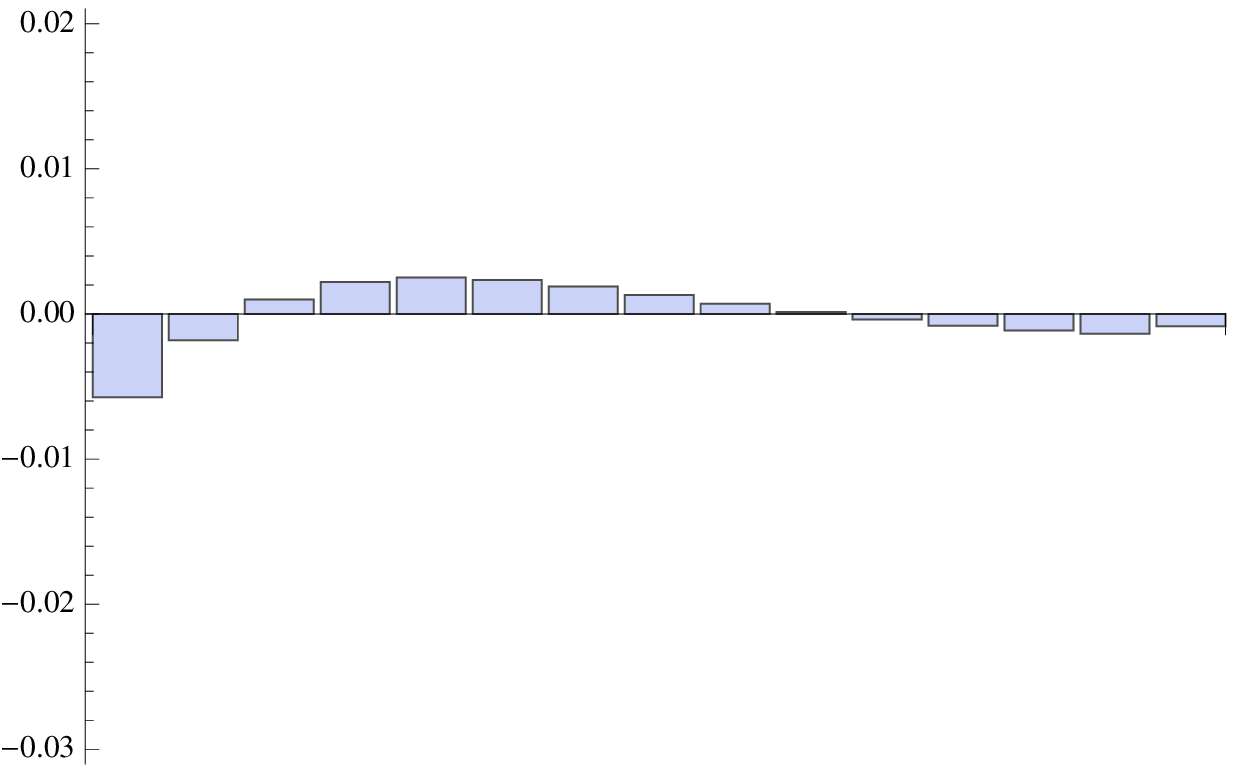}
 \hskip0.7cm
\includegraphics[width=4.75cm,height=1.5cm] {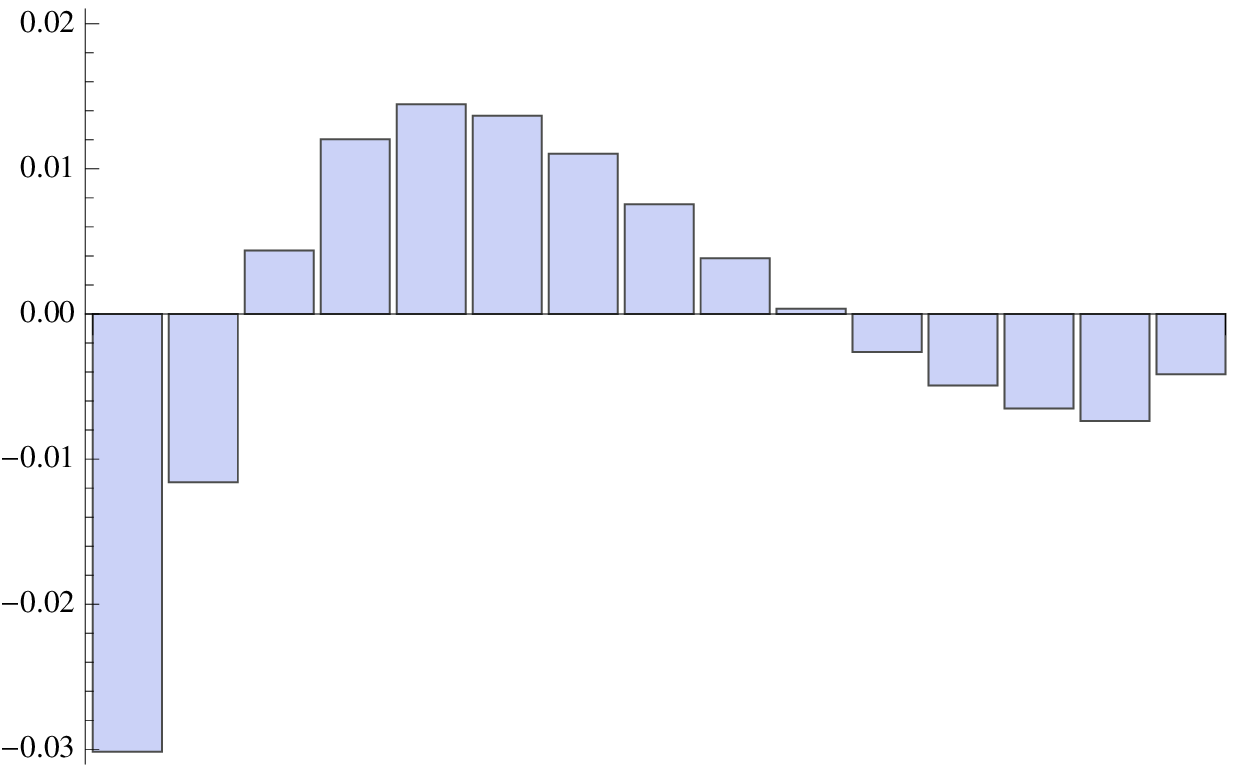}
\end{center}
\caption{Asymptotic solutions evolving from the same initial data as in the Example $1$.
In the first row, on the left [resp., on the right] is the asymptotic stationary distribution of the model in which the welfare is given by the formula
$(\ref{formula welfare})$ with $\gamma = 0.45$ [resp., with $\gamma = 0.15$]; 
The histograms in the second row (scaled differently w.r. to those in the first one) depict 
the difference between the fraction of individuals in each class pertaining to two models: these are:
on the left [resp., on the right], the model with $\gamma = 0.45$ [resp., with $\gamma = 0.15$] and that one 
with the same welfare for all classes.}
\vskip0.75cm
\begin{center}
\includegraphics[width=4.75cm,height=2.5cm] {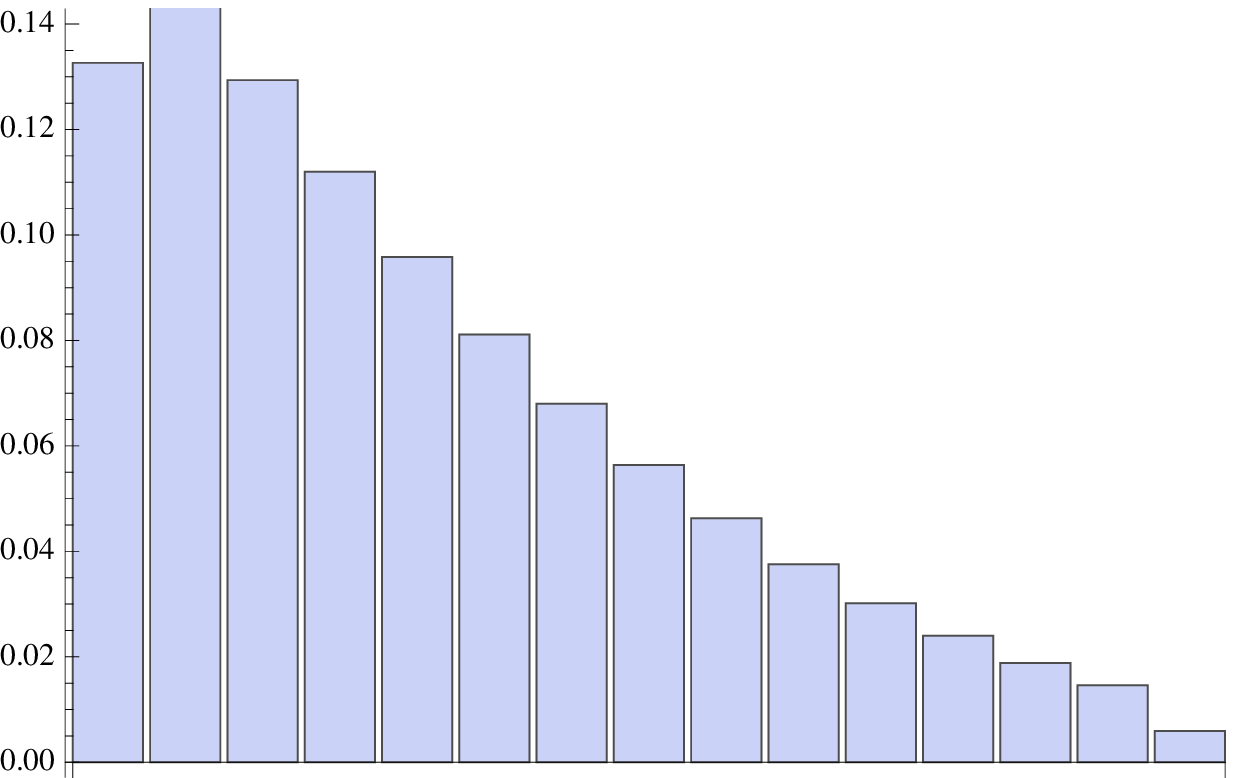}
 \hskip0.7cm
\includegraphics[width=4.75cm,height=2.5cm] {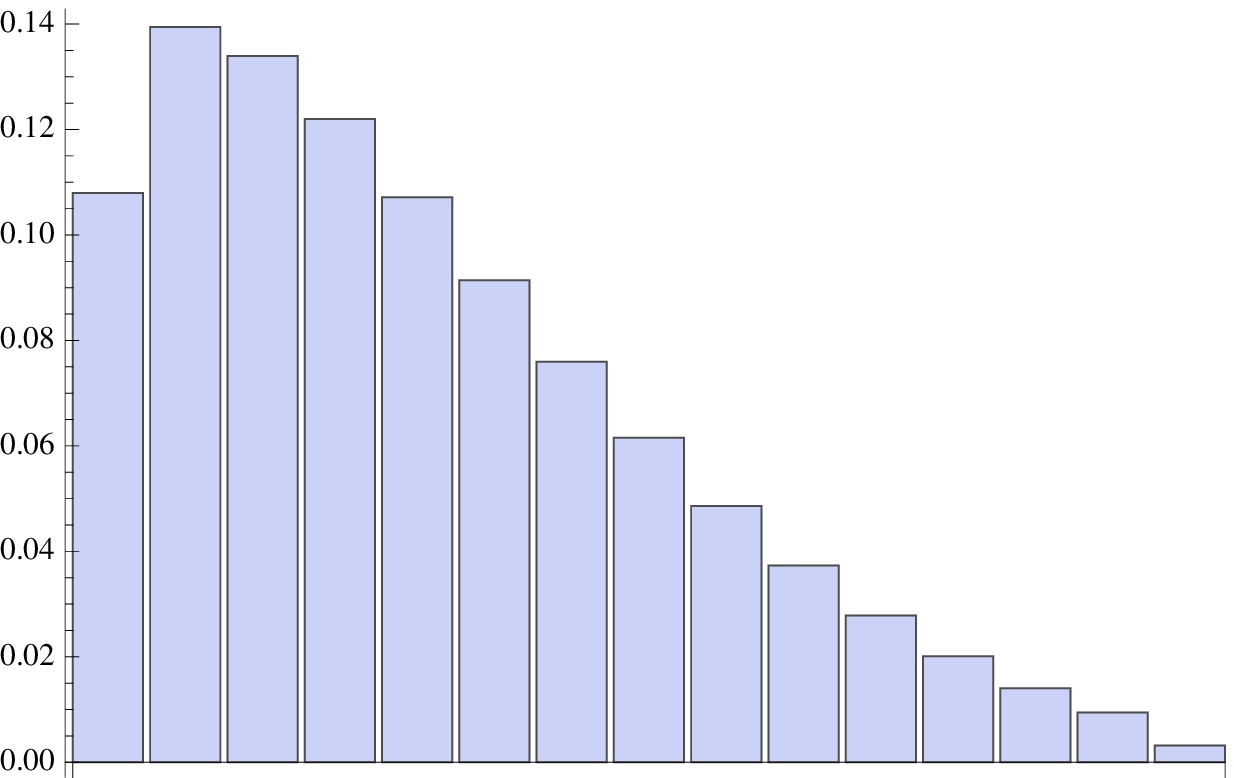}
\end{center}
\begin{center}
\includegraphics[width=4.75cm,height=1.55cm] {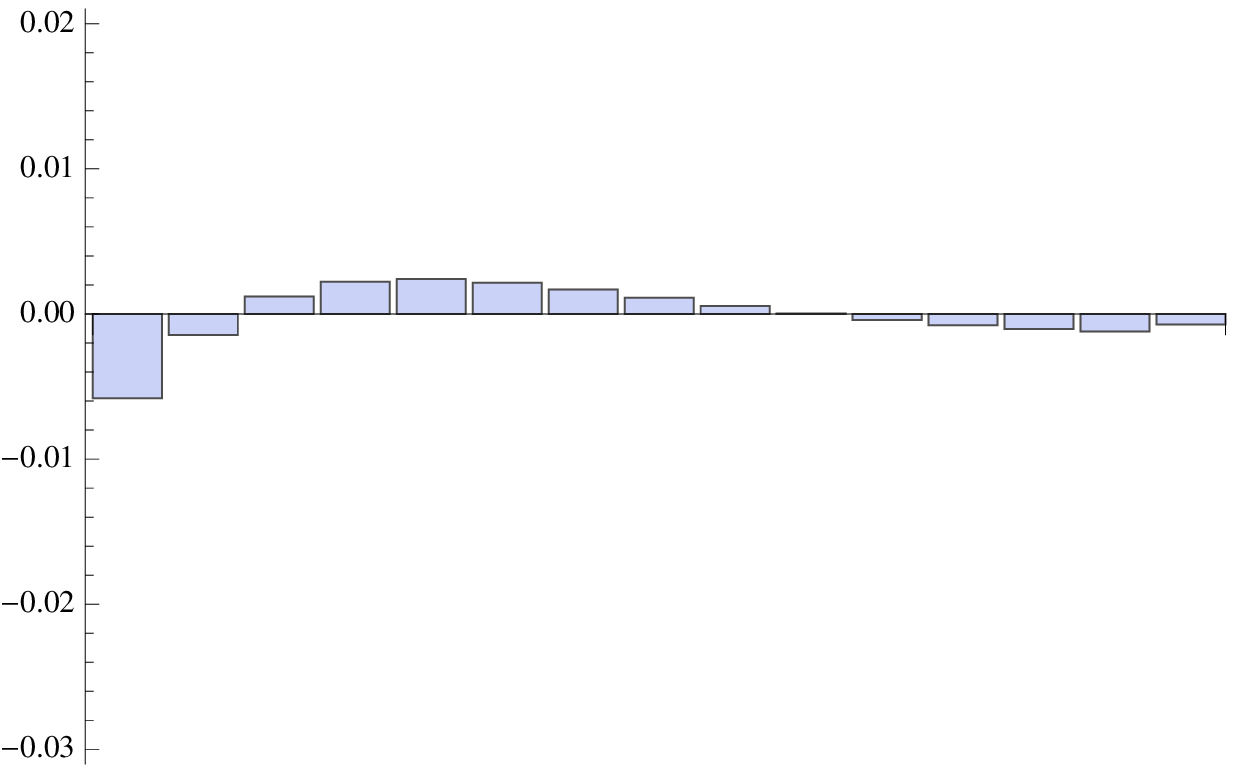}
 \hskip0.7cm
\includegraphics[width=4.75cm,height=1.5cm] {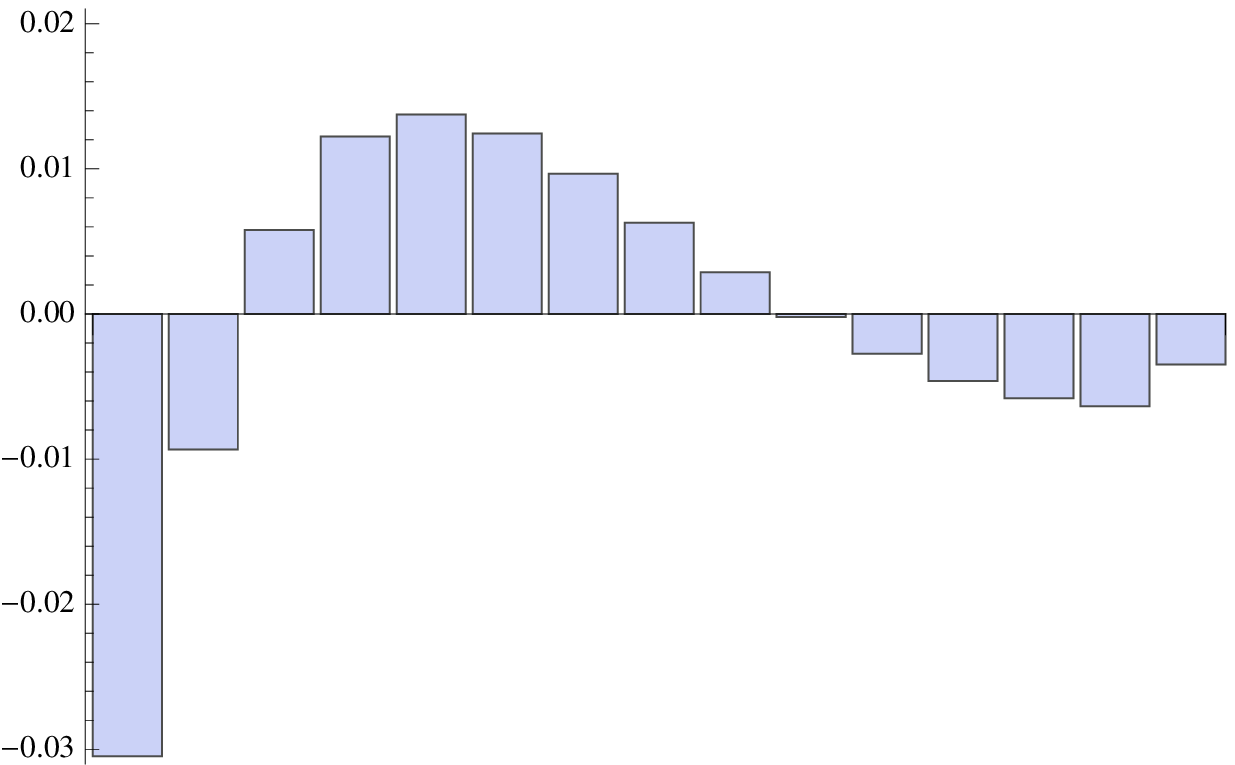}
\end{center}
\caption{Asymptotic solutions evolving from the same initial data as in the Example $2$.
In the first row, on the left [resp., on the right] is the asymptotic stationary distribution of the model in which the welfare is given by the formula
$(\ref{formula welfare})$ with $\gamma = 0.45$ [resp., with $\gamma = 0.15$]; 
The histograms in the second row (scaled differently w.r. to those in the first one) depict 
the difference between the fraction of individuals in each class pertaining to two models: these are:
on the left [resp., on the right], the model with $\gamma = 0.45$ [resp., with $\gamma = 0.15$] and that one 
with the same welfare for all classes.}
\end{figure*}

\medskip

The following can be concluded based on these and similar simulations. 

\smallskip

\noindent {\bf {Dependence of the asymptotic stationary distribution on differently weighted welfare measures}}.
The profile of the asymptotic stationary distribution depends on $\gamma$.
When $\gamma$ decreases and all other data are kept unchanged,
at the asymptotic equilibrium an increase of the fraction of individuals belonging to the middle classes 
(and, correspondingly, a decrease of those in the poorest and richest classes) can be detected.

\subsection{Some comparisons and observations}
\begin{figure*}[h]
\begin{center}
\includegraphics[width=5.25cm,height=3.0cm] {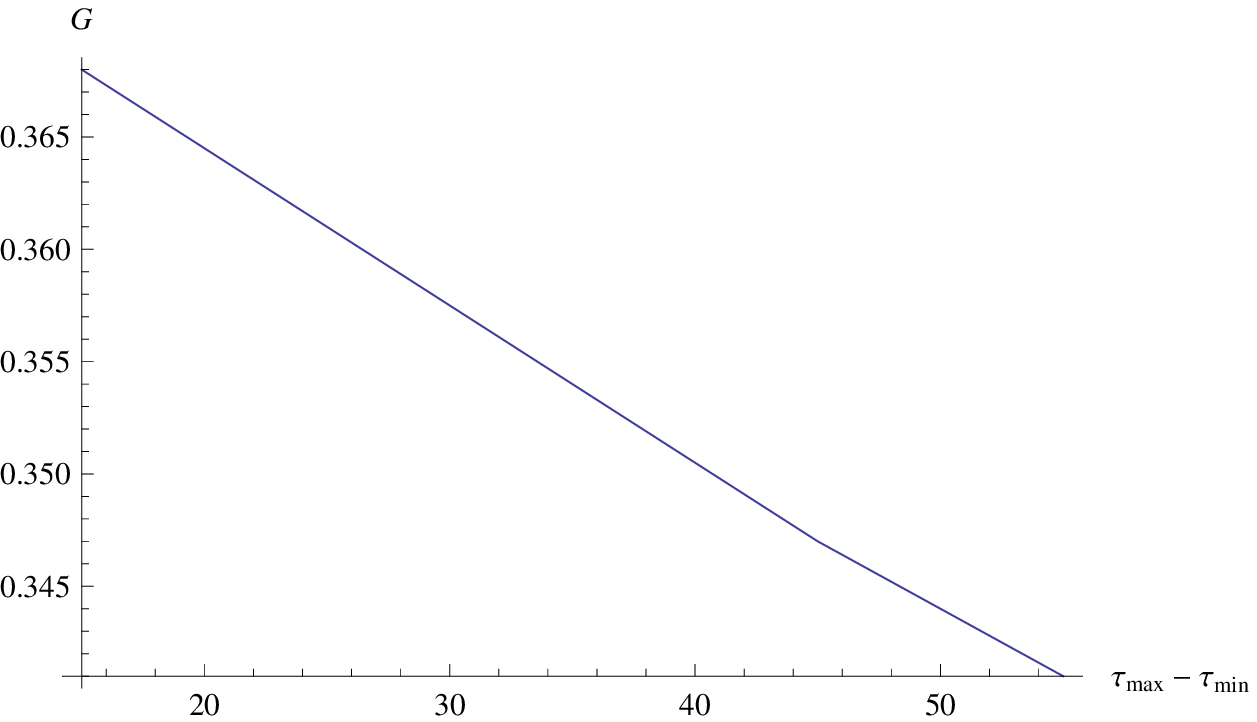}
 \hskip0.7cm
\includegraphics[width=5.25cm,height=3.0cm] {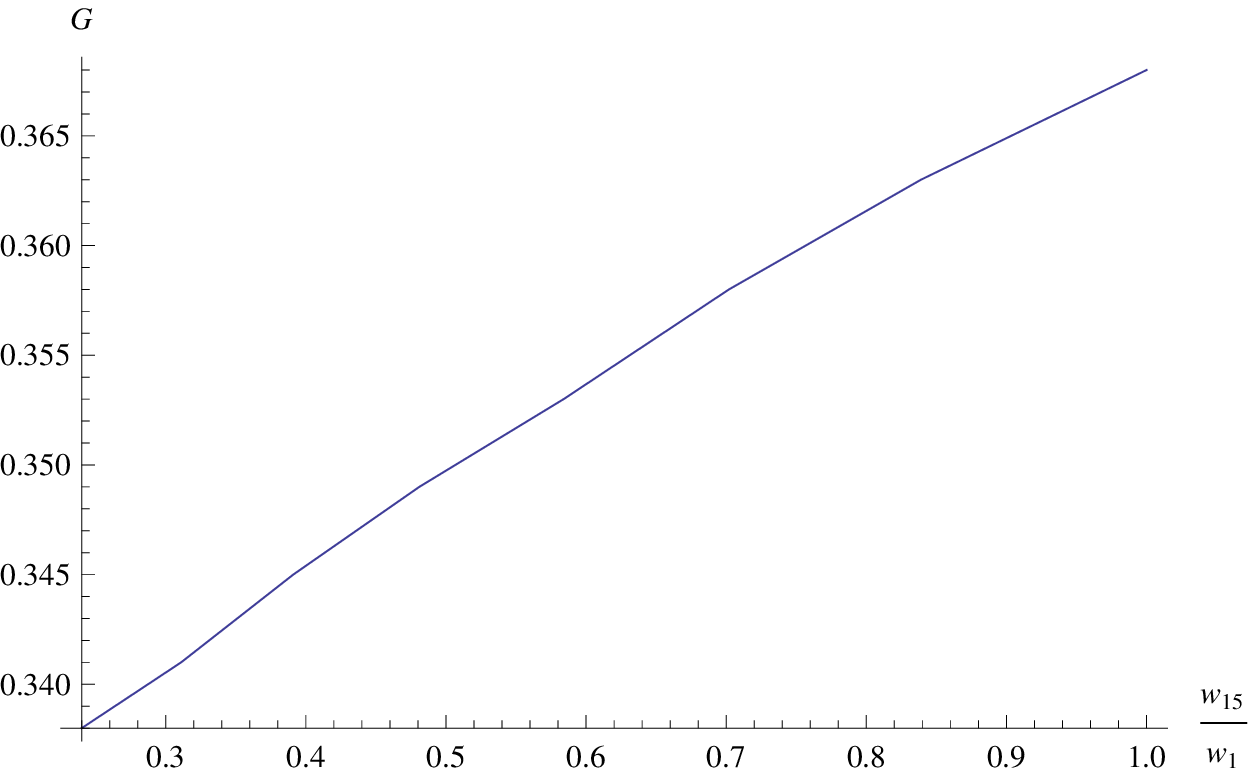}
\end{center}
\caption{Left:  dependence of the inequality Gini index $G$ on the difference $\Delta \tau=(\tau_{max}-\tau_{min})$
between
the maximum and the minimum tax rate (for the richest and poorest class respectively).
The Gini index decreases almost linearly when $\Delta \tau$ increases.
Right: dependence of $G$ on the
ratio $w_{15}/w_1$ between the welfare provision assigned to an individual
of the income class $15$ and $1$. $G$ decreases almost linearly when the ratio decreases.
These two panels refer to the data and to the linear fits in the Tables $\rm{\ref{tab:1varyingtaxrates}}$ and $\rm{\ref{tab:1diffwelfare}}$.}
\label{fig:graphsoftheTables1and3} 
\end{figure*}

The data of Table $\rm{\ref{tab:1varyingtaxrates}}$ show that the dependence of the Gini index $G$ on the difference $\Delta \tau=(\tau_{max}-\tau_{min})$ 
is almost exactly linear, see Figure \ref{fig:graphsoftheTables1and3}, left panel.
The regression line has equation $G = -0.0007 \Delta \tau + 0.378$ ($R^2=0.9991$) and this can be expressed by saying that 
in order to obtain a 1\% reduction of the Gini index, $\Delta \tau$ must increase by approx. 15\%. This
should be compared with the dependence of $G$ on changes in welfare redistribution. For instance, a linear fit of the data of Table
$\rm{\ref{tab:1diffwelfare}}$ gives a regression line
$G= 0.04 (w_{15}/w_1)+0.3291$ ($R^2=0.9955$), see Figure \ref{fig:graphsoftheTables1and3}, right panel. 
This means that a 1\% reduction of $G$
can be obtained approx.\ by a 25\% diminution of the ratio $w_{15}/w_1$, 
which appears to be a far less ``invasive'' policy measure 
(it is a 25\% cut in the welfare received from the richest class, compared to that received from the poorest; 
the intermediate classes undergo proportional cuts).

Finally, the data of Tables 
$\rm{\ref{tab:1varyingtaxrates}}$-$\rm{\ref{tab:2diffwelfare}}$
show a remarkable decrease in the total tax revenue when the gap $\Delta \tau$ is increased, 
while the contrary occurs when the welfare for the rich (the ratio $w_{15}/w_1$) is decreased. 
This means that in our model the inequality-reduction policy of ``taxing the rich much more than the poor'' is 
not only less efficient than ``cutting the welfare for the rich'', but also leads to a reduction in the government budget and thus possibly to a loss of jobs.

\section{Concluding remarks}
Our models are based on a system of differential equations of the kinetic discretized-Boltzmann kind. 
Society is described as an ensemble of individuals divided into a finite number of income classes; 
these individuals exchange money through binary and ternary interactions, leaving the total wealth unchanged. 
The interactions occur with a certain predefined frequency and several other parameters can also be set, 
in order to provide a probabilistic representation 
as realistic as possible. For instance, we can fix the probability that in an encounter between two individuals 
the one who pays is the rich or the poor; we can make the exchanged amount depend on the income classes 
(variable saving propensity), etc.
After a sufficiently long time the solution of the equations reaches an equilibrium state characterized by an income distribution, 
which depends on the total income and on the interaction parameters, but not on the initial distribution.
We emphasize here that it would be possible to introduce further heterogeneity in
various meaningful ways. For instance, one could assign specific interaction frequencies to the income classes 
through additional coefficients to be inserted in the equations,
or introduce a network structure 
or ``behavioral sectors'', which comprise honest taxpayers, occasional tax evaders, etc. (this is the object of work in preparation, \cite{Bertotti M.L. Modanese G. 5}).

The ternary interactions represent 
taxation and redistribution effects: they express the subtraction,
in correspondence to each binary transfer, 
of an amount whose percentage (tax rate) depends on the income classes of the individuals involved in the interaction;
and they define the redistribution of this amount to all other individuals.
In the simplest version of the model the redistribution is uniform. In order to represent a more realistic welfare system 
we have introduced in this work a weighted redistribution: the poorest classes receive a larger part of the total tax revenue, 
according to a linearly increasing function. This may describe a means-tested welfare system or 
a policy of limitation of the ``indirect subsidies for the rich''.  
We have then investigated the dependence on this redistribution parameter of the inequality of the society, 
as measured from the Gini index or from the shape of the income distribution.

The Gini inequality indices $G$ of the income profiles of our model turn out to be quite realistic. Actually, we can easily compare ``pre-redistribution" values of G 
(those obtained when taxation terms are switched off) with the values after redistribution. Detailed real data for such quantities have recently become available \cite{Solt F.}. 
With a pre-redistribution value of $G$ of ca. $0.46$ for the numerical solutions of Table $\rm{\ref{tab:2diffwelfare}}$, and a post-redistribution value of $G$ which varies
between $0.34$ and $0.38$, depending on taxation rates and welfare parameters, it turns out that we are quite close to the real data of the 
United States (while, for instance, economies like Germany and Denmark exhibit a markedly larger
redistribution gap).

Still concerning the redistribution aspect, we would like to stress that this aspect is present also in Asset Exchange Models based on purely physical analogies, but only the present paper proposes, to our knowledge, a
redistribution which can be tuned on the single income classes, thus simulating the working principle of a real means-tested welfare policy. In other words, physical analogies can represent well situations where, for instance,
particles dissipate energy through radiation, and the radiation is re-distributed to the whole system; but it is not possible to tune this redistribution in an arbitrary way.

A further advantage of our model is to allow the evaluation of social mobility \cite{Bertotti M.L. Modanese G. 5}. This cannot be extracted, of course, 
from the equilibrium income distribution alone, but requires consideration of the probabilities of class
advancement. Our numerical results concerning social mobility confirm the empirical correlation observed between mobility and equality \cite{Corak M.,WilPic}. 
Using the tunable welfare scheme introduced in this paper, it has even been possible to find numerically some ``equi-Gini lines" in the tax-rates/welfare plane $\tau$-$w$, 
i.e. to locate in the plane of these two variables some continuous lines along which the variation of tax rates and welfare parameters leave the inequality index constant.

In conclusion, we found that in order to diminish inequality, 
a policy of reduction of the welfare and subsidies for the rich classes is more effective than an increase in the taxation rate gap $\tau_{max}-\tau_{min}$. 
This same policy also has the effect of slightly increasing the total tax revenue, instead of decreasing it.
It therefore avoids potentially painful cuts in the government budget. 
The results obtained from these simplified models have clearly a limited validity, but we believe that they
are conceptually interesting 
and can serve as a basis and stimulus for further analysis.



\bibliographystyle{ws-acs}
\bibliography{ws-acs}

\end{document}